\documentclass[%
reprint,
 amsmath,amssymb,
 aps,
 pre,
]{revtex4-2}

\usepackage{graphicx} 
\usepackage{dcolumn} 
\usepackage{bm}          
\usepackage{hyperref}  
\usepackage{comment} 
\usepackage{soul}
\usepackage{xurl}

\hypersetup{
  colorlinks=true,
  urlcolor=blue,
  linkcolor=blue,
  citecolor=blue,
  filecolor=blue,
  breaklinks=true
}

\DeclareMathOperator\arccosh{arccosh}

\newcommand{\Leq}{L_{\text{eq}}}
\newcommand{\req}{r_{\text{eq}}}
\newcommand{\rmax}{r_{\text{max}}}
\newcommand{\xhl}{x_{r}^{\text{hl}}}
\newcommand{\Lres}{L_{\text{res}}}
\newcommand{\tauc}{\tau_{\mathrm{cap}}}
\newcommand*\diff{\mathop{}\!\mathrm{d}}

\newcommand{\rev}[1]{{\color{black} #1}}

\begin{document}

\preprint{APS/123-QED}

\title{\rev{Resetting-induced instability in queues fed by a search process in an interval}}

\author{Jos\'{e} Giral-Barajas}
\email{j.giral-barajas24@imperial.ac.uk}
\author{Paul C. Bressloff}
\email{p.bressloff@imperial.ac.uk}
\affiliation{%
 Department of Mathematics, Imperial College London, London SW7 2AZ, United Kingdom
}%


\begin{abstract}
Proper management of resources whose arrival and consumption are subject to environmental randomness is an intrinsic process in both natural and artificial systems. This phenomenon can be modeled as a queuing process whose arrival distribution is determined by a search process with stochastic resetting. When the queuing system has a limited number of servers and the search process occurs within a bounded domain, the dynamics of expediting or delaying the search through stochastic resetting interact with the long-term dynamics of the number of resources in the queue. We combine results from queuing theory with those from search processes with stochastic resetting in a bounded domain to obtain regions of the parameter space of the search process that ensure convergence of the number of resources in the queue to a steady state. Furthermore, we find a threshold resetting rate at which the effects of stochastic resetting shift from reducing convergence regions to expanding them. Finally, we demonstrate that this threshold value grows exponentially with the number of servers, making it harder for stochastic resetting to improve the convergence of the queueing system.
\end{abstract}

                              
\maketitle

\section{\label{sec:Intro}Introduction}
A topic of major interest is the time it takes a particle diffusing within a bounded domain $\Omega$ to find a specific target $\mathcal{U}\subseteq\Omega$ \cite{grebenkov_target_2024}. Examples include biochemical reaction kinetics \cite{weiss_first_1967, benichou_geometry-controlled_2010, kumar_inference_2023, barman_fluctuations_2025}, cellular transport \cite{bressloff_stochastic_2013, iyer-biswas_first-passage_2016, bressloff_modeling_2020, stotsky_random_2021}, and animal foraging \cite{fauchald_using_2003, benichou_optimal_2005, viswanathan_levy_2008, bidari_stochastic_2022}. This phenomenon is frequently modeled as a Brownian motion in $\Omega$ with a totally absorbing boundary in the inner boundary $\partial\mathcal{U}$ and a reflecting boundary in the outer boundary $\partial\Omega$. In scenarios like this, in which $\Omega$ is bounded, the mean time to find the target is finite, and the behavior of the \textit{first-passage time} (FPT) is well understood \cite{redner_guide_2001}.

A common assumption made in studies concerning FPTs is that the process terminates once the searcher finds the target $\mathcal{U}$ \cite{redner_applications_2014}. This can be thought of as a \textit{searcher-centric model}. However, there are several examples of search processes in which the searcher delivers cargo upon finding the target, then continues, allowing the searcher to return to its initial position, reload cargo, and restart the search. In the long term, repeating these steps leads to the accumulation of resources within the target. Therefore, this scenario can be thought of as a \textit{target-centric model}.

\begin{figure}[ht!]
\includegraphics[width=\columnwidth]{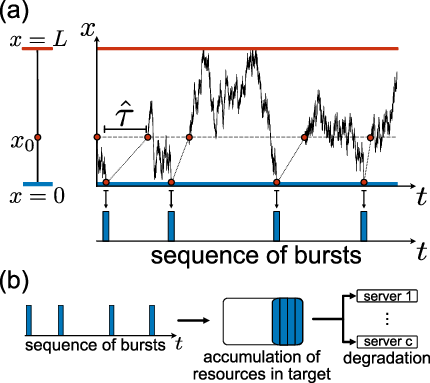}
\caption{\label{fig:SaC}Sequence of search-and-capture events for a diffusing particle in the interval with an absorbing boundary at the origin and a reflecting boundary at the endpoint mapped into a $G/M/c$ queueing process. (a) The particle follows a one-dimensional (1D) Brownian motion starting from $x_{0}\in[0,L]$ and searching for the target at $x=0$. Once the particle finds the target, it delivers a resource, returns to its starting position, and restarts the process. In each round, the delivery, return, and reload take a random time $\hat{\tau}$. Repeating this process produces a sequence of delivery times, known as the sequence of bursts. (b) Under the additional assumption that the random consumption times are exponentially distributed, the sequence of bursts can be fed into a $G/M/c$ queueing process to track the accumulation of resources within the target.}
\end{figure}

The target-centric model has been explored recently by allowing a search process to return to its initial position to reload and restart the search after finding the target \cite{bressloff_search-and-capture_2019}. The total time it takes the particle to unload the cargo, return to the initial position, and reload is described by the \textit{refractory period} $\hat{\tau}$. This refractory period can be treated as a random variable with probability density function $\varphi(\hat{\tau})$ and mean $\tauc$. This is illustrated in Fig.~\ref{fig:SaC}~(a). The repetition of these steps is commonly known as a search-and-capture process and has been explored in different scenarios \cite{bressloff_search-and-capture_2019, bressloff_target_2020, bressloff_first-passage_2021}.

The search-and-capture process produces a sequence of delivery times, which we refer to as burst events; see Fig.~\ref{fig:SaC}~(a). As time progresses, the  \rev{cumulative} number of bursts increases, and resources begin to accumulate at the target. \rev{To counterbalance this buildup, we assume that the target consumes the accumulated resources via a fixed number of consumption channels. These dynamics can be mapped onto a queueing process \cite{bressloff_queueing_2020, giral-barajas_stochastic_2025}. In this analogy, each burst event is identified with the arrival of a customer, while resource consumption is identified with service completion in the queue. The target itself plays the role of the queueing system, and its consumption channels play the role of servers, each removing resources after a random service (consumption) time \cite{bhat_introduction_2015}. Consequently, the number of accumulated resources at the target at any given time is equivalent to the number of customers in the queueing system; see Fig.~\ref{fig:SaC}~(b). This mapping allows us to describe resource accumulation in terms of a queue whose arrival process is determined by the search-and-capture dynamics and whose service process is determined by the consumption times.}

A queueing system is determined by its maximum capacity, the queue discipline---referring to the order in which customers receive their service---, the customers' arrival distribution, the customers' service distribution, and the number of servers \cite{allen_probability_1978}. Under the additional assumptions that the target has an infinite capacity and a \textit{first-come, first-served} (FCFS) discipline, a queue process can be described using Kendall's shorthand notation $A/B/c$, where $A$ describes the inter-arrival time distribution, $B$ the service time distribution, and $c$ the number of servers \cite{kendall_stochastic_1953}. When the accumulated resources are being consumed by a finite number, $c<\infty$, of service stations with exponential service time distribution---i.e., Markovian service times---, the number of resources within a target being delivered by a search-and-capture process can be modeled with a $G/M/c$ system \cite{giral-barajas_stochastic_2025}. Here $G$ denotes a general (non-Markovian) inter-arrival time distribution, $\mathcal{F}_{0}(t)$, determined by the search-and-capture process, and $M$ denotes Markovian service times. This mapping allows us to study the long-term behavior of the number of resources in the target by leveraging the wide range of results from classical queuing theory, in particular those from the $G/M/c$ system. 

It is known that there exists a steady-state number of customers in the $G/M/c$ system if and only if the customer arrival rate, $\lambda$, is lower than the total customer service rate, $c\mu$, where $\mu$ is the service rate per server \cite{takacs_introduction_1962}. This condition can be rewritten as $\rho<1$ where $\rho=\lambda/(c\mu)$ is known as the \textit{traffic intensity} \cite{cooper_introduction_1981}. In a recent work, we utilized this result to determine zones of convergence to a steady-state number of customers in the parameter space of the search process in a \textit{one-dimensional} (1D) bounded interval \cite{giral-barajas_stochastic_2025}. In particular, for fixed interval length, diffusivity, starting position and service rate there exists a crossover initial position, $x_{0}^{*}$, for which the number of resources in the $G/M/c$ system associated with the target converges to a steady state if and only if $x_{0}>x_{0}^{*}$. Therefore, starting the process closer than $x_{0}^{*}$ to the target is detrimental for the long-term stability of the number of resources.

Another topic of current interest is search processes with stochastic resetting \cite{evans_stochastic_2020}. A typical motivating example is a diffusing particle searching for a specific target $\mathcal{U}$ in an unbounded domain $\Omega$. The introduction of stochastic resetting, in which the particle is allowed to reset to a given position $x_{r}\in\Omega$ following a sequence of random times with resetting rate $r$, renders the \textit{mean first passage time} (MFPT) finite \cite{evans_diffusion_2011}. Moreover, the MFPT can be optimized in terms of the resetting rate \cite{evans_diffusion_2011-1, evans_diffusion_2014}. There are several stochastic processes showing this behavior, including L\'{e}vy flights \cite{kusmierz_first_2014, kusmierz_optimal_2015}, run-and-tumble particles \cite{evans_run_2018}, processes with resetting in a potential landscape \cite{pal_diffusion_2015, ray_diffusion_2020, roberts_ratchet-mediated_2024}, and processes in bounded domains \cite{durang_first-passage_2019}.

It is well understood that resetting can expedite the search process and optimize the MFPT as a function of the resetting rate  \cite{evans_diffusion_2011}. This fact is particularly relevant for search processes in unbounded domains whose behavior is null recurrent, as is the case for diffusion processes in unbounded domains \cite{evans_diffusion_2011-1, evans_diffusion_2014}. However, scenarios in which the process is intrinsically positive recurrent, and the MFPT is finite without the need for resetting, require further analysis to determine when resetting expedites the search process \cite{christou_diffusion_2015, pal_landau-like_2019, pal_inspection_2022, mendez_nonstandard_2022}. 

\rev{This is the case with search processes in bounded domains, making them processes with rich dynamics. Therefore, stochastic resetting in bounded domains continues to be investigated in a variety of settings. Recent studies have extended the basic framework to more realistic models, including search in spatially heterogeneous lattice environments \cite{barbini_lattice_2024}, targets with random position \cite{evans_stochastic_2025}, spatiotemporally dependent resetting rates and combinations of bulk and boundary resetting \cite{garcia-valladares_how_2026}, as well as threshold-induced resetting \cite{de_bruyne_optimization_2020} and its extension to systems of $N$ noninteracting searchers \cite{biswas_target_2025}. In the present work, however, we restrict attention to the basic case of a diffusing particle in a 1D interval.}

The particular scenario of a diffusing particle in a 1D interval with two absorbing boundaries has been thoroughly explored in Refs.~\cite{pal_first_2019,villarroel_double_2022}. \rev{Ref.~\cite{pal_first_2019}} introduces a threshold subregion of the interval $[0,L]$ defined as $\mathcal{D}:=(0,x_{0-})\cup(x_{0+},L)$ where $x_{0\pm}=(5\pm\sqrt{5})/10$, such that if $x_{r}\in\mathcal{D}$ resetting is capable of expediting the search process and if $x_{r}\notin\mathcal{D}$ resetting will always increase the unconditional MFPT. In the searcher-centric approach, this acceleration of the search process is usually interpreted as beneficial. However, in the target-centric approach, expediting the search process and reducing the MFPT can push the number of resources in the target out of equilibrium, break convergence to a steady state, and shift the process dynamics toward unbounded growth.

The main goal of this paper is to explore how stochastic resetting affects convergence to the steady-state distribution of the number of resources in a $G/M/c$ queueing system arising from a search-and-capture process driven by a diffusing particle in a 1D bounded interval. To do so, we combine results for the threshold subset of initial positions allowing expediting of the search process by the introduction of stochastic resetting, $\mathcal{D}$, from Ref.~\cite{pal_first_2019} and results for the threshold convergence-inducing starting position, $x_{0}^{*}$, from Ref.~\cite{giral-barajas_stochastic_2025} to derive new regions of convergence in the parameter space.

This paper is organized as follows. In Section~\ref{sec:2} we introduce the modelling rationale and present previous results of first-passage processes, queuing theory, and stochastic resetting, which will serve as building blocks for the main results of the paper. In Section~\ref{sec:3}, we extend the results for the regions of convergence in the parameter plane $(L,x_{0})$ and for the critical starting position to consider the effects of stochastic resetting while fixing the resetting rate. We analyze the effects of resetting and find a threshold interval length, $\Leq$, for which the inclusion of resetting at the predefined fixed resetting rate changes the convergence zone from widening to shrinking. Finally, in Section~\ref{sec:4}, we analyze the critical regions for a fixed interval length while varying the resetting rate. We derive conditions that ensure the region of starting positions for the search process allowing resetting to expedite the process intersects the region of starting positions that ensure convergence of the queueing system to a steady state, and explore this scenario. We find a threshold resetting rate, $\req$, for which the effects of resetting change from shrinking the convergence zone to widening it, and a resetting rate, $\rmax$, that maximizes the critical, convergence-inducing starting position and minimizes the length of the convergence zone. We conclude by analyzing the behavior of $\req$ and $\rmax$ as the number of servers in the $G/M/c$ queueing system increases.
\section{\label{sec:2}Search-and-capture, queueing theory and resetting in the interval}
In this section, we present previous results on first-passage processes, queueing theory, and stochastic resetting and establish the modelling rationale. We then build on these results in subsequent sections.

\subsection{\label{subsec:2A}Search-and-capture events in an interval without resetting}
Consider a searching particle subject to a Brownian motion inside a 1D bounded domain $[0,L]$ with an absorbing target at 0 and a reflecting boundary at $L$, see Fig.~\ref{fig:SaC}~(a). We denote by $X_{t}$ the particle's position at time $t$ and by $p(x,t|x_{0})$ the probability density function for the particle to be in $x$ at time $t$, having started at $x_{0}$. This probability density function evolves according to the diffusion equation
\begin{equation}
	\frac{\partial p(x,t|x_{0})}{\partial t} = D\frac{\partial^{2}p(x,t|x_{0})}{\partial x^{2}}=-\frac{\partial J(x,t|x_{0})}{\partial x}\;,
	\label{eq:DiffusionEquation}
\end{equation}
where $D$ is the diffusivity and $J(x,t|x_{0}):=-D\partial_{x}p(x,t|x_{0})$ is the \textit{probability flux}. This diffusion equation is supplemented with an absorbing boundary condition at 0, $p(0,t|x_{0})=0$, a reflecting boundary condition at $L$, $\partial_{x}p(L,t|x_{0})=0$, and the initial condition $p(x,0|x_{0})=\delta(x-x_{0})$.

Define the FPT to be absorbed by the target at 0 as
\begin{equation}
	\mathcal{T}_{0}(x_{0}) := \inf\{t>0 : X_{t}=0,\;X_{0}=x_{0}\}\;,
	\label{eq:FPTDefinition}
\end{equation}
As there is only one absorbing target, the FPT's probability density function, denoted by $f_{0}(x_{0},t)$, is equivalent to the probability flux into the target, denoted by $J_{0}(x_{0},t)$,
\begin{equation}
	f_{0}(x_{0},t) = J_{0}(x_{0},t) = D\frac{\partial p(x,t|x_{0})}{\partial x}\Big|_{x=0}\;.
	\label{eq:FPTDensity}
\end{equation}
Hereafter, we will use the probability flux into the target and the FPT's probability density function interchangeably.

We can now compute the probability that the particle eventually gets absorbed by the target, commonly called the \textit{splitting probability}, as
\begin{equation}
	\pi_{0}(x_{0}) := \mathbb{P}[\mathcal{T}_{0}(x_{0})<\infty]=\int_{0}^{\infty}J_{0}(x_{0},t')\diff t'=\widetilde{J}_{0}(x_{0},0)\;,
	\label{eq:SplittingProbability}
\end{equation}
where $\widetilde{J}_{0}(x_{0},s)$ denotes the Laplace transform of the probability flux into the boundary, defined by
\[\widetilde{J}_{0}(x_{0},s) = \int_{0}^{\infty} e^{-st}J_{0}(x_{0},t)\diff t\;.\]

Additionally, the MFPT can be described in terms of the probability flux in Laplace space. From Eq.~(\ref{eq:FPTDensity}) we directly have that $\widetilde{f}_{0}(x_{0},s) = \widetilde{J}_{0}(x_{0},s)$, from where we obtain that
\begin{equation}
	T_{0}(x_{0}) := \mathbb{E}[\mathcal{T}_{0}(x_{0})]=-\frac{\partial}{\partial s}\widetilde{J}_{0}(x_{0},s)\Big|_{s=0}\;.
	\label{eq:MFPTLaplace}
\end{equation}
Therefore, all the quantities of interest for the search process can be described via the Laplace transform of the probability flux into the target, $\widetilde{J}_{0}(x_{0},s)$.

An explicit expression for $\widetilde{J}_{0}(x_{0},s)$ can be found using the Green's function for a diffusing particle inside an interval with an absorbing boundary at the origin and a reflecting boundary at the endpoint \cite{redner_guide_2001}
\begin{equation}
	\widetilde{J}_{0}(x_{0},s)=\frac{\cosh\left[\sqrt{\frac{s}{D}}(L-x_{0})\right]}{\cosh{\left[\sqrt{\frac{s}{D}}L\right]}}\;.
	\label{eq:LaplaceFlux}
\end{equation}
Combining Eq.~(\ref{eq:LaplaceFlux}) with Eqs.~(\ref{eq:SplittingProbability}-\ref{eq:MFPTLaplace}), we obtain explicit expressions for the splitting probability and the MFPT

\begin{subequations}
\begin{eqnarray}
\pi_{0}(x_{0})&=&1\;, \label{eq:Splitting}
\\
T_{0}(x_{0})&=&\frac{(2L-x_{0})x_{0}}{2D}\;. \label{eq:MFPT}
\end{eqnarray}
\end{subequations}

As mentioned in the introduction, each time the searcher finds the target, it delivers a resource and, instead of terminating the process, it returns to its starting position to reload cargo and start a new search. This shifts the process's focus from the searcher to the target, accumulating the resources delivered in each burst event.

The behavior of the burst events is completely determined by their inter-arrival times. Given that the search-and-capture cycles are independent, the sequence of inter-arrival times is a sequence of independent, identically distributed random variables (iidrv). We denote the $n$th inter-arrival time by $\Delta_{n}$ and have that
\begin{equation}
	\Delta_{n} = \tau_{n}-\tau_{n-1} \stackrel{d}{=} \hat{\tau} + \mathcal{T}(x_{0})\;,\;
	\label{eq:IATs}
\end{equation}
where $\tau_{n}$ denotes the time of the $n$th burst event and $\Delta_{n} \stackrel{d}{=} \hat{\tau} + \mathcal{T}(x_{0})$ denotes equality in distribution. From Eq.~(\ref{eq:IATs}), we have that the mean inter-arrival time is given by $T_{0}(x_{0})+\tauc$ and the mean bursting rate is $\lambda = (T_{0}(x_{0})+\tauc)^{-1}$. Moreover, the inter-arrival time density, denoted by $\mathcal{F}_{0}$, can be directly computed as the convolution of the FPT density and the refractory period's density. Dropping the dependance on $x_{0}$ from the notation, we have that
\begin{equation}
	\mathcal{F}_{0}(t) = \int_{0}^{t}f_{0}(t')\varphi(t-t')\diff t'\;.
	\label{eq:IATDensity}
\end{equation}
Laplace transforming Eq.~(\ref{eq:IATDensity}) yields
\begin{equation}
	\widetilde{\mathcal{F}}_{0}(s) = \widetilde{f}_{0}(s)\widetilde{\varphi}(s)\;.
	\label{eq:IATDensityLaplace}
\end{equation}
\begin{figure*}[t]
\includegraphics[width=0.9\textwidth]{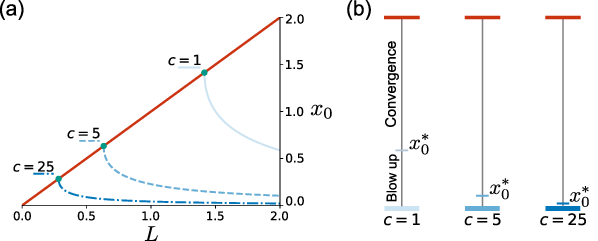}
\caption{\label{fig:CriticalRegions}Critical spatial configurations for the existence of a steady-state distribution with instantaneous refractory periods. (a) The threshold starting position of the search process ensures that the number of resources in the $G/M/c$ system converges to a steady state. Each curve is determined by Eq.~(\ref{eq:CriticalX0}) as a function of $L$, fixing $D=1$, $\mu=1$, and $\tauc=0$ and varying the number of servers in the system. Green dots represent the threshold interval length, $L^{*}$, determined by Eq.~(\ref{eq:CriticalL}). (b) Critical initial position and convergence and blow up zones for an interval of length $L=2$ and different numbers of servers. This representation identifies with the endpoints of the curves in panel (a), when $L=2$. Whenever $x_{0}>x_{0}^{*}$, the number of resources converges to a steady state, and whenever $x_{0}\leq x_{0}^{*}$, the number of resources blows up.}
\end{figure*}

\subsection{\label{subsec:2B}Convergence zones for the G/M/c without resetting}
As highlighted in the introduction, the accumulation of resources after several rounds of the search-and-capture process can be mapped onto a queueing process. To do so, we need to determine the departure process to counterbalance the continual arrival coming from the search-and-capture process. Heuristically, two scenarios counteract accumulation: degradation and consumption. In the case of degradation, each resource begins service independently of the others upon its arrival. In the case of consumption, where the system has a finite number of servers to consume resources, the start of service for each resource is subject to server availability---determined by the number of resources present in the system upon arrival.

Assuming service times follow an exponential distribution, these two scenarios correspond to distinct types of queuing processes. When resources are subject to degradation, the accumulation can be modeled with a $G/M/\infty$ queue. When the target consumes resources, the accumulation can be modeled with a $G/M/c$ queue. In both cases, $G$ represents a general (non-Markov) inter-arrival time distribution, and $M$ represents a Markovian service time distribution. In the first case, $\infty$ denotes an infinite number of servers, while in the second case, there is a finite number of servers, $c$. These models behave differently over long periods. While the number of resources in the $G/M/\infty$ system always converges to a steady state, the convergence of the $G/M/c$ system depends on traffic intensity. Thus, the number of resources in the $G/M/c$ system converges to a steady state if and only if
\begin{equation}
	\rho := \frac{\lambda}{c\mu} < 1\;,
	\label{eq:TrafficIntensity}
\end{equation}
where $\lambda^{-1}$ is the mean inter-arrival time and $\mu^{-1}$ is the mean service time \cite{cooper_introduction_1981}. Note that the mean inter-arrival time for the burst events generated by the search and capture process is given by $T_{0}(x_{0})+\tauc$, where $T_{0}(x_{0})$ is the MFPT of the search process, given by Eq.~(\ref{eq:MFPT}) for our setting of a 1D Brownian motion within an interval with a single absorbing target at the origin.

In a recent work, we studied the convergence of a $G/M/c$ queue emerging from a search-and-capture process in a bounded interval. In particular, we combined the condition for convergence in Eq.~(\ref{eq:TrafficIntensity}) with the explicit expression of the MFPT in Eq.~(\ref{eq:MFPT}) to derive the following condition for convergence to a steady-state number of resources \cite{giral-barajas_stochastic_2025}: $x_{0}>x_{0}^{*}$ where 
\begin{equation}
	x_{0}^{*}=L-\sqrt{L^{2}-\frac{2D}{c\mu}+2D\tauc}\;.
	\label{eq:CriticalX0}
\end{equation}
We also obtained a minimal interval length that ensures the existence of regions in the search process's parameter space that allow the G/M/c system to converge to a steady state:
\begin{equation}
	L^{*}=\sqrt{\frac{2D}{c\mu}-2D\tauc}\;.
	\label{eq:CriticalL}
\end{equation}
To do so, we equated $x_{0}^{*}$, given by Eq.~(\ref{eq:CriticalX0}), to $L$ and solved the resulting equation for $L$. This crossover interval length can be interpreted as follows. If the search process occurs in an interval of length $L\leq L^{*}$, with $L^{*}$ determined by fixed values of $D$, $\mu$, $\tauc$, and $c$, the number of resources accumulated in the target at $0$ will blow up regardless of the initial position, see Fig.~\ref{fig:CriticalRegions}~(a). On the contrary, if $L>L^{*}$, there exists a subregion of the interval, given by $(x_{0}^{*},L]$, such that if the search process starts in said region, the number of resources in the target converges to a steady state, see Fig.~\ref{fig:CriticalRegions}~(b)
\subsection{\label{subsec:2C}Stochastic resetting in the interval}
\begin{figure}[h!]
\includegraphics[width=\columnwidth]{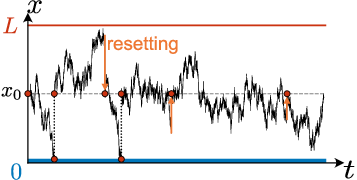}
\caption{\label{fig:SaC_Res}Sample path of a search-and-capture process with instantaneous resetting and instantaneous refractory periods. The resetting position is assumed to be the same as the initial position, $x_{0}=x_{r}$. The dynamics introduced by stochastic resetting are shown in orange straight arrows, and the restart dynamics of the search-and-capture process are shown in dotted lines.}
\end{figure}

Now suppose that in each round of the search-and-capture process, the particle is allowed to instantaneously reset with rate $r$ to its starting position before delivering its current cargo to the target, according to a sequence of independent random times, each following the same distribution $\psi(t)$, see Fig.~\ref{fig:SaC_Res}. Stochastic resetting has been explored using renewal theory for non-instantaneous protocols and general resetting distributions \cite{rotbart_michaelis-menten_2015, belan_restart_2018, chechkin_random_2018, bressloff_search_2020, bodrova_resetting_2020}. However, for the sake of concreteness, we limit this work to the scenario of instantaneous, exponential resetting, i.e. $\psi(t)=re^{-rt}$.

The Laplace transform of the FPT density and the MFPT have been derived for the instantaneous exponential resetting scenario \cite{bressloff_queueing_2020, reuveni_optimal_2016}. Denoting by $f_{r}(x_{0},t)$ de FPT density with resetting and by $T_{r}(x_{0})$ the MFPT density with resetting, we have that
\begin{equation}
	\widetilde{f}_{r}(x_{0},s) = \frac{(r+s)\widetilde{f}_{0}(x_{0},r+s)}{s+r\widetilde{f}_{0}(x_{0},r+s)}\;,
	\label{eq:FPTdensity_resetting}
\end{equation}
and
\begin{equation}
	T_{r}(x_{0}) = \frac{1-\widetilde{f}_{0}(x_{0},r)}{r\widetilde{f}_{0}(x_{0},r)}\;.
	\label{eq:MFPT_resetting_0}
\end{equation}
Substituting the explicit expression for $\widetilde{f}_{0}$ in Eq.~(\ref{eq:LaplaceFlux}) into the expression for the MFPT time we obtain
\begin{equation}
	T_{r}(x_{0}) = \frac{1}{r}\left[\frac{\cosh\left[\sqrt{\frac{r}{D}}L\right]}{\cosh{\left[\sqrt{\frac{r}{D}}(L-x_{0})\right]}}-1\right]\;.
	\label{eq:MFPT_resetting}
\end{equation}

As mentioned in the introduction, when the search process is intrinsically positive recurrent and the MFPT is finite, the inclusion of stochastic resetting can delay the process and increase the MFPT. This is the case for a diffusing particle in a 1D interval with an absorbing boundary at the origin. To determine whether or not the MFPT with resetting can be improved and has at least one critical point, one needs to analyze the sign of the derivative $\diff T_{r}/\diff r$ at $r=0$ \cite{pal_first_2019,reuveni_optimal_2016, pal_search_2020}. When the derivative is negative, stochastic resetting can expedite the search process in the small-$r$ regime. This condition is equivalent to
\begin{equation}
	\frac{\sigma_{0}(x_{0})}{T_{0}(x_{0})}>1\;,
	\label{eq:CVCondition}
\end{equation}
where $\sigma_{0}^{2}(x_{0})=\text{Var}(\mathcal{T}_{0}(x_{0}))$ \cite{pal_first_2017}. It can be shown that for 1D Brownian motion in the interval $[0,L]$ with an absorbing boundary at the origin, the inequality in Eq.~(\ref{eq:CVCondition}) is fulfilled if and only if
\begin{equation}
	x_{0}\in\mathcal{X}_{0}:=\left(0,\left[1-\frac{1}{\sqrt{5}}\right]L\right)\;.
	\label{eq:CriticalSubregion}
\end{equation}
For the complete computations, see Appendix~\ref{app:A}. Therefore, as long as $x_{0}\in\mathcal{X}_{0}$, the MFPT as a function of the resetting rate is non-monotonic and initially decreasing. This ensures that there exists a subset of resetting rates $r>0$ for which resetting expedites the search process. Moreover, there exists a unique minimizer $r_{\rm opt}$ for the MFPT.

\begin{figure*}[t!]
\includegraphics[width=\textwidth]{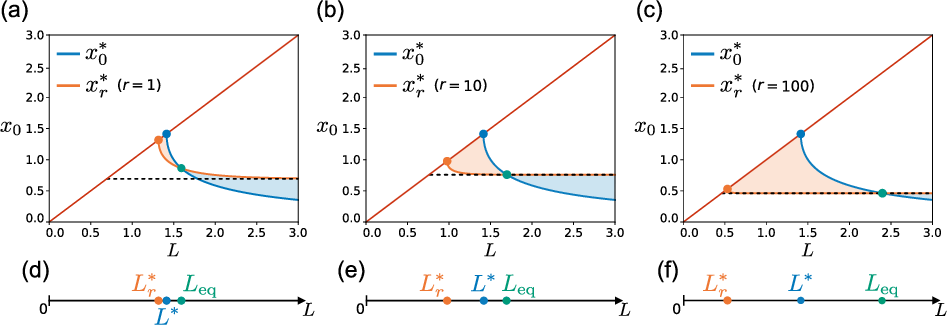}
\caption{\label{fig:critical_resetting}Critical, convergence-inducing initial conditions without stochastic resetting and with fixed resetting rates of (a) $r=1$, (b) $r=10$, and (c) $r=100$. The \rev{blue curves} correspond to the scenario without stochastic resetting, determined by Eq.~(\ref{eq:CriticalX0}), and the \rev{orange curves} correspond to the scenario with stochastic resetting, determined by Eq.~(\ref{eq:CriticalStartingResetting}). The \rev{orange-shaded} area represents the \rev{additional} resetting-induced spatial configurations that ensure convergence, compared to the case without resetting, while the \rev{blue-shaded} area represents the lost spatial configurations that ensure convergence when stochastic resetting is introduced. The dotted \rev{black} line corresponds to the critical starting position in the half-line, $\xhl$, shown in Eq.~(\ref{eq:CriticalLimitResetting}). Panels (d-f) show the threshold quantities $L_{r}^{*}$, $L^{*}$, and $\Leq$ corresponding to the panel directly above. In every case, we fix $D=1$, $\tauc=0$, $\mu=1$, and $c=1$.}
\end{figure*}
\section{\label{sec:3}Critical regions for a fixed resetting rate}
We begin our analysis by fixing a resetting rate $r>0$ and deriving convergence conditions for the $G/M/c$ queue associated with the target at the origin. Hereafter, we fix the units of time and length by setting the consumption rate $\mu=1$ and diffusivity $D=1$. As we have an explicit expression for the MFPT with stochastic resetting given by Eq.~(\ref{eq:MFPT_resetting}), the analysis can be done following the procedure in Ref.~\cite{giral-barajas_stochastic_2025} regardless of whether the stochastic resetting speeds up the search process or slows it down. The inter-arrival times have the same distribution as the sum of the FPT, denoted $\mathcal{T}_{r}$ for the process with stochastic resetting, and the refractory period. This implies that the inter-arrival time density of the burst events, denoted as $\mathcal{F}_{r}$ for the process with stochastic resetting, satisfies $\widetilde{\mathcal{F}}_{r}(s)=\widetilde{f}_{r}(s)\widetilde{\varphi}(s)$ and the mean inter-arrival time is given by $T_{r}(x_{0})+\tauc$. We now use the explicit expression for the MFPT in Eq.~(\ref{eq:MFPT_resetting}) to compute the arrival rate as
\begin{equation}
	\lambda_{r} = \frac{r\cosh{\left[\sqrt{\frac{r}{D}}(L-x_{0})\right]}}{\cosh\left[\sqrt{\frac{r}{D}}L\right]+(r\tauc-1)\cosh{\left[\sqrt{\frac{r}{D}}(L-x_{0})\right]}}\;.
	\label{eq:ArrivalRate}
\end{equation}

Substituting this expression into the convergence condition in Eq.~(\ref{eq:TrafficIntensity}) and solving the inequality for the initial position, $x_{0}$, we obtain the following critical starting position
\begin{equation}
	x_{r}^{*}=L-\sqrt{\frac{D}{r}}\arccosh\left[{\frac{c\mu\cosh{\left[\sqrt{\frac{r}{D}}L\right]}}{r+c\mu-rc\mu\tauc}}\right]\;.
	\label{eq:CriticalStartingResetting}
\end{equation}
Therefore, for a fixed resetting rate $r>0$, whenever the search process starts further than $x_{r}^{*}$, i.e. $x_{0}>x_{r}^{*}$, the traffic intensity at the $G/M/c$ queue is lower than one, and the number of resources in the system converges to a steady state. However, not all interval lengths allow for convergence. For each fixed number of servers, $c>0$, there are values of $L$ for which $x_{r}^{*}>L$ and the number of resources in the system blows up regardless of the rest of the parameters of the search process. To find the minimal convergence-allowing interval length, denoted by $L_{r}^{*}$, we equate $x_{r}^{*}$ to $L$ in Eq.~(\ref{eq:CriticalStartingResetting}) and solve the resulting equation for $L$. We obtain that
\begin{equation}
	L_{r}^{*}=\sqrt{\frac{D}{r}}\arccosh\left[\frac{r+c\mu-rc\mu\tauc}{c\mu}\right]\;.
	\label{eq:CriticalLengthResetting}
\end{equation}
Consequently, for each fixed number of servers and resetting rate, $L>L_{r}^{*}$ and $x_{0}>x_{r}^{*}$ ensure that the number of resources in the system converges to a steady state. This condition splits the parameter space for the spatial configurations of the search process into two regions, see Fig.~\ref{fig:critical_resetting}~(a)-(c).

It can be shown that $\lim_{r\to0^{+}}L_{r}^{*}=L^{*}$ and that $L_{r}^{*}$ is a monotonically decreasing function of $r$. Therefore, for each $r>0$ we have that $L_{r}^{*}<L^{*}$. This enlarges the set of interval lengths that allow convergence to a steady state for the $G/M/c$, see Fig.~\ref{fig:critical_resetting}~(d)-(f). The introduction of resetting also changes the limit behavior of the critical starting position. First, observe that in the case without resetting $x_{0}^{*}$, seen as a function of the number of servers and the interval length, is such that $\lim_{c\to\infty}x_{0}^{*}=0$ and $\lim_{L\to\infty}x_{0}^{*}=0$. The former limit makes the traffic intensity zero, allowing any spatial configuration for the search process without breaking convergence. The latter coincides with the fact that a Brownian motion in the half line has an infinite MFPT to the target at the origin, again making the traffic intensity zero and allowing any spatial configuration for the search process without breaking convergence. In contrast, the case with resetting is such that $\lim_{c\to\infty}x_{r}^{*}=0$ and
\begin{equation}
	\lim_{L\to\infty}x_{r}^{*}=\sqrt{\frac{D}{r}}\log\left(\frac{r+c\mu-rc\mu\tauc}{c\mu}\right)=:\xhl\;,
	\label{eq:CriticalLimitResetting}
\end{equation}
where $\xhl$ denotes the critical, convergence-inducing starting position in the half line. Again, the former limit makes the traffic intensity zero, allowing any spatial configuration for the search process without breaking convergence. However, now the latter limit coincides with the fact that a Brownian motion with stochastic resetting in the half line has a finite MFPT given by
\begin{equation}
	T_{r}(x_{0})=\frac{1}{r}(e^{\sqrt{\frac{r}{D}}x_{0}}-1)\;,
	\label{eq:MFPTHalfLine}
\end{equation}
as shown in \cite{evans_diffusion_2011, evans_diffusion_2011-1}, see Fig.~\ref{fig:critical_resetting}~(a)-(c). Therefore, in the presence of stochastic resetting, the only way to ensure convergence to a steady-state number of resources in the $G/M/c$ queue, regardless of the parameters of the search process, is to increase the number of servers.

To conclude this section, we explore the differences in the convergence and blow-up regions in the parameter space as a function of the interval length, with and without resetting. The key thing to note is that for any given resetting rate $r>0$, there exists an interval length such that the critical starting position with resetting is the same as the one without resetting. Denoting this interval length by $\Leq$, and seeing $x_{r}^{*}$ and $x_{0}^{*}$ as functions of $L$, this is $x_{r}^{*}(\Leq) = x_{0}^{*}(\Leq)$. In principle, this interval length can be found by equating Eq.~(\ref{eq:CriticalX0}) to Eq.~(\ref{eq:CriticalStartingResetting}) and solving for $L$. However, the resulting expression has no analytical solution, and $\Leq$ is determined by the following implicit expression
\begin{widetext}
\begin{equation}
	(r+c\mu+rc\mu\tauc)\cosh\left[\sqrt{\frac{r\Leq^{2}}{D}-\frac{2r}{c\mu}+2r\tauc}\right]=c\mu\cosh\left[\sqrt{\frac{r}{D}}\Leq\right]\;.
	\label{eq:Leq}
\end{equation}
\end{widetext}

This value marks a transition in the effects of stochastic resetting within the regions of convergence for the $G/M/c$ queue. For $L\in[L_{r}^*,\Leq)$, the presence of stochastic resetting widens the convergence zone. It introduces new possible spatial configurations for the search process that allow the number of resources in the queueing system to converge, see  \rev{orange-shaded} areas in Fig.~\ref{fig:critical_resetting}~(a)-(c). In contrast, for $L\in(\Leq,\infty)$ the presence of stochastic resetting shrinks the convergence zone and reduces the spatial configurations for the search process that allow convergence of the number of resources in the queueing system, see \rev{blue-shaded} areas in Fig.~\ref{fig:critical_resetting}~(a)-(c).

Observing Fig.~\ref{fig:critical_resetting}~(a)-(c), we also realize that as the resetting rate increases the convergence stated in Eq.~(\ref{eq:CriticalLimitResetting}) seems to happen faster; with $x_{r}^{*}$ being almost indistinguishable from $\xhl$ for small values of $L$. If we assume that $x_{r}^{*}=\xhl$, we can find an explicit expression for $\Leq$ by equating Eq.~(\ref{eq:CriticalX0}) to Eq.~(\ref{eq:CriticalLimitResetting}) and solving for $L$, obtaining
\begin{equation}
	\Leq=\sqrt{\frac{D}{r}}\frac{2r+2rc\mu\tauc+ c\mu\log\left(\frac{r+c\mu-rc\mu\tauc}{c\mu}\right)^{2}}{2c\mu\log\left(\frac{r+c\mu-rc\mu\tauc}{c\mu}\right)}\;.
	\label{eq:LeqScaled}
\end{equation}
To determine the validity of this explicit expression, we explore the rate of convergence of $x_{r}^{*}$ to $\xhl$, as $L\to\infty$, for different values of the resetting rate. Let $r>0$, and define $d_{r}(L):=x_{r}^{*}(L)-\xhl$. Numerical computations of the limit of $d_{r}(L)e^{2\sqrt{\frac{r}{D}}L}$ as $L\to\infty$ show that, as long as $\tauc<(c\mu)^{-1}+r^{-1}$,
\begin{equation}
	d_{r}(L)\stackrel{L}{\sim}\sqrt{\frac{D}{r}}\frac{r(c\mu\tauc-1)[r(c\mu\tauc-1)-2c\mu]}{(c\mu)^{2}}e^{-2\sqrt{\frac{r}{D}}L}\;,
	\label{eq:LeqScaledError}
\end{equation}
where $f\stackrel{L}{\sim}g$ denotes that $f(L)/g(L)\to1$ as $L\to\infty$. In particular, in the scenario of instantaneous refractory periods, i.e., $\tauc=0$, for all $r>0$ we have that
\begin{equation}
	d_{r}(L)\stackrel{L}{\sim}\sqrt{\frac{D}{r}}\frac{r(r+2c\mu)}{(c\mu)^{2}}e^{-2\sqrt{\frac{r}{D}}L}\;.
	\label{eq:LeqScaledError2}
\end{equation}
These results are yet to be analytically shown; however, they show a fast decay in the error, coinciding with the behavior observed in Fig.~\ref{fig:critical_resetting}~(a)-(c). Moreover, they establish the accuracy of the explicit expression for $\Leq$ in Eq.~(\ref{eq:LeqScaled}) for finite, small values of $r$ and $L$, and describe its error.
\section{\label{sec:4}Critical subregions for a fixed interval length}
We now shift the focus to the effects of varying the resetting rate while holding the interval length fixed. In this setting, we have two crossover quantities for the initial position of the search process. First, we have the threshold initial condition $x_{0}^{*}$ in the absence of resetting, for which if $x_{0}>x_{0}^{*}$ then the traffic intensity in the $G/M/c$ queue is lower than one---allowing the number of resources in the system to converge to a steady state. We also have the subregion $\mathcal{X}_{0}=(0,(1-1/\sqrt{5})L)$, derived in Subsection~\ref{subsec:2C} following Ref.~\cite{pal_first_2019}, such that if $x_{0}\in\mathcal{X}_{0}$ then stochastic resetting can expedite the search process, thus reducing the MFPT and increasing the traffic intensity in the $G/M/c$ queue. 

\begin{figure}[t!]
\includegraphics[width=\columnwidth]{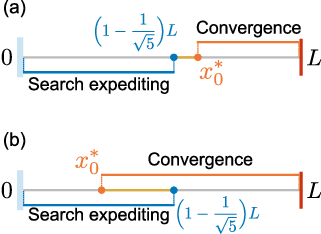}
\caption{\label{fig:ZoneOfInterest}Different scenarios of interest for the exploration of the effects of stochastic resetting in the long-term behavior of the $G/M/c$ queueing system. (a) When $x_{0}^{*}>(1-1/\sqrt{5})L$, the zone of interest for the initial position of the search process is $((1-1/\sqrt{5})L,x_{0}^{*})$. (b) When $x_{0}^{*}<(1-1/\sqrt{5})L$, the zone of interest for the initial position of the search process is $(x_{0}^{*},(1-1/\sqrt{5})L)$.}
\end{figure}

There are two possible scenarios involving these two crossover initial positions. In the first scenario, when $x_{0}^{*}>(1-1/\sqrt{5})L$, the subregion $\mathcal{X}_{0}$ is a subset of the initial positions of the search process that make the number of resources in the $G/M/c$ queue blow up. Hence, if the initial position of the search process is such that $(1-1/\sqrt{5})L<x_{0}<x_{0}^{*}$ (see Fig.~\ref{fig:ZoneOfInterest}~(a)) then the introduction of stochastic resetting hinders the search process. This implies that increasing $r$ results in $x_0$ lying to the right of the critical point $x_r^*$ and the long-term dynamics of the $G/M/c$ queue are shifted from blow up to convergence. In the second scenario,  when $x_{0}^{*}<(1-1/\sqrt{5})L$, the subregion $\mathcal{X}_{0}$ intersects the zone of convergence-inducing starting positions. Therefore, if $x_{0}^{*}<x_{0}<(1-1/\sqrt{5})L$ (see Fig.~\ref{fig:ZoneOfInterest}~(b)), there exists a possibility that stochastic resetting shifts the long-term dynamics from convergence to blow up. This occurs if $x_r^*$ shifts to the right of $x_0^*$ and $x_0$.

The first scenario is less interesting, since it is always true that, with a sufficiently high resetting rate, one can shift the long-term dynamics of the queue from blow up to convergence. This is given by the fact that $\lim_{r\to\infty}T_{r}(x_{0})=\infty$, and one can make the traffic intensity as small as required by incrementing the resetting rate. However, in the second scenario, it is not immediately clear when stochastic resetting can cause the number of resources in the queue to blow up. This issue is the focus of the remainder of Section~\ref{sec:4}.

We begin by determining the conditions ensuring that $x_{0}^{*}<(1-1/\sqrt{5})L$. To do so, we substitute the expression for $x_{0}^{*}$ given in Eq.~(\ref{eq:CriticalX0}) into the inequality and solve for $L$. We obtain that $x_{0}^{*}<(1-1/\sqrt{5})L$ if and only if
\begin{equation}
	L>\Lres:=\sqrt{\frac{5D}{2}\left(\frac{1}{c\mu}-\tauc\right)}\;.
	\label{eq:LZone}
\end{equation}
Note that this condition is only well-defined for $\tauc\leq(c\mu)^{-1}$, and whenever $\tauc>(c\mu)^{-1}$ we have that $x_{0}^{*}=0$ for all $L>0$. This is since the condition for convergence to steady state in terms of the traffic intensity in Eq.~(\ref{eq:TrafficIntensity}) can be rewritten as
\begin{equation}
	\frac{1}{c\mu}<T_{r}(x_{0})+\tauc\;.
	\label{eq:TrafficIntensity2}
\end{equation}
Therefore, as $T_{r}(x_{0})\geq0$, whenever $\tauc>(c\mu)^{-1}$ the condition for convergence is automatically fulfilled regardless of the spatial configuration of the search process. Hereafter, we will assume that $\tauc\leq(c\mu)^{-1}$ and $L>\Lres\geq0$.

\begin{figure}[b!]
\includegraphics[width=\columnwidth]{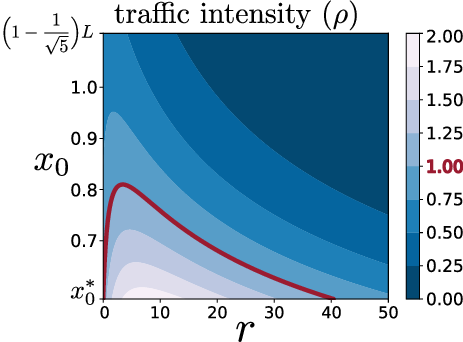}
\caption{\label{fig:TrafficIntensity}Contour plots for the traffic intensity in the $(r,x_{0})$ plane. The starting position is conditioned to the zone of interest $(x_{0}^{*},(1-1/\sqrt{5})L)$. The critical curve defining the phase transition between convergence and blow-up for the number of resources in the queue is shown in red. The rest of the parameters of the model are $D=1$, $L=2$, $\tauc=0$, $\mu=1$, and $c=1$.}
\end{figure}

If the initial position of the search process lies in the range $x_{0}^{*}<x_{0}<(1-1/\sqrt{5})L$, then stochastic resetting can expedite the search. That is, the MFPT $T_r(x_0)$ will be a non-monotonic function of the resetting rate with a unique minimizing rate $r_{\rm opt}$ for each value of $x_{0}$ \cite{reuveni_optimal_2016, pal_search_2020}. Additionally, the MFPT is an increasing function of the resetting position, which also corresponds to the starting position of the search process. On the other hand, traffic intensity is inversely proportional to the MFPT. Therefore, it is a decreasing function of the initial position, and for each fixed $x_{0}$, it has a unique maximizing resetting rate equal to $r_{\rm opt}$. 

\begin{figure}[b!]
\includegraphics[width=\columnwidth]{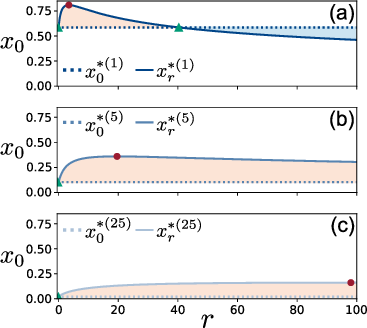}
\caption{\label{fig:CriticalRegionsConvergence}Critical, convergence-inducing starting position with and without resetting for different numbers of servers in the $G/M/c$ queueing system. Plots of $x_{0}^{*}$ and $x_{r}^{*}$ for 1 server in (a), 5 servers in (b), and 25 servers in (c). The \rev{orange-shaded} area represents the resetting-induced shrinkage of the convergence zone in the $(r,x_{0})$ plane, and the  \rev{blue-shaded} area represents the resetting-induced expansion of the convergence zone in the $(r,x_{0})$ plane. The red dots represent the maximum $x_{r}^{*}$ at $r_{\max}$ and the green triangles represent the resetting rate, $\req$, for which $x_{r}^{*}=x_{0}^{*}$. The rest of the parameters are $D=1$, $L=2$, $\tauc=0$, and $\mu=1$.}
\end{figure}

We explore the combination of these dynamics for the particular case of the $G/M/1$ queue in Fig.~\ref{fig:TrafficIntensity}. We observe that the subset $(0,\infty)\times[x_{0}^{*},(1-1/\sqrt{5})L]$ of the parameter plane $(r,x_{0})$ contains a subregion in which the long-term behavior of the number of resources in the queueing system shifts from convergence to blow-up. Furthermore, the threshold between convergence and blow up can be determined by the critical starting position $x_{r}^{*}$, given in Eq.~(\ref{eq:CriticalStartingResetting}), as a function of $r$. That is, for each resetting rate $r>0$, as long as $x_{0}>x_{r}^{*}$, the number of resources in the system will still converge to a steady-state distribution. However, non-monotonicity of the MFPT with respect to $r$ implies that varying $r$ can shift $x_r^*$ to the right of $x_0^*$ (and thus possibly $x_0$), resulting in blow up. This is illustrated in Fig.~\ref{fig:CriticalRegionsConvergence} for different numbers of servers $c$. It can be seen that in each case there exists a unique resetting rate $r=r_{\max}(c)$ that maximizes the critical, convergence-inducing starting position $x_r^*$ and thus minimizes the convergence zone. Note that $r_{\max}(c)$ can be found by setting $\frac{\partial}{\partial r}x_{r}^{*} = 0$ and numerically solving the resulting equation for $r$. The latter is given in Appendix~\ref{app:B}. In Fig.~\ref{fig:CriticalRegionsConvergence}, we indicate the numerical solution for different numbers of servers as red dots overlayed in the critical curves $x_{r}^{*}$.

Fig.~\ref{fig:CriticalRegionsConvergence} also establishes that, for each fixed number of servers, there is a crossover point $\req(c)$ at which $x_0^*=x_r^*$. As with $\Leq$, the resetting rate $\req$ is obtained by numerically solving the implicit Eq.~(\ref{eq:Leq}). Example values are indicated in Fig.~\ref{fig:CriticalRegionsConvergence} as green triangles overlaid on the critical curves $x_{r}^{*}$. The critical resetting rate $\req$ marks a transition in the effects of stochastic resetting within the regions of convergence for the $G/M/c$ queue. For $r\in(0,\req)$, the presence of stochastic resetting shrinks the convergence zone and reduces the spatial configurations for the search process that allow convergence of the number of resources in the queueing system. This shrinkage is maximized at $\rmax\in(0,\req)$, producing the smallest convergence zone for the queueing system. In contrast, for $r\in(\req,\infty)$, the presence of stochastic resetting widens the convergence zone. It introduces new possible spatial configurations for the search process that allow the number of resources in the queueing system to converge. Ultimately, as $\lim_{r\to\infty}x_{r}^{*} = 0$, the blow-up zone can be reduced as much as needed for a big enough $r>\req$.

\begin{figure}[b!]
\includegraphics[width=\columnwidth]{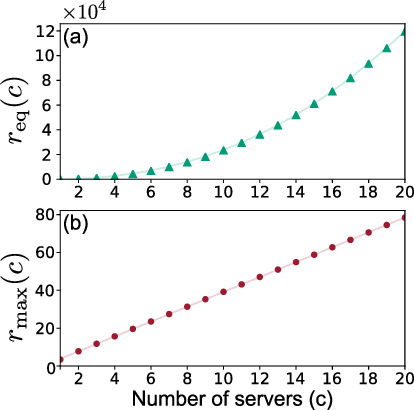}
\caption{\label{fig:ReqRmax}Maximizing and equalizing resetting rates for $x_{r}^{*}$ as a function of the number of servers. (a) corresponds to the values of $\req(c)$, displayed as green triangles, and (b) corresponds to the values of $\rmax(c)$, displayed as red dots. The rest of the parameters are $D=1$, $\mu=1$, $L=2$, and $\tauc=0$.}
\end{figure}

To conclude this section, we explore the behavior of $r_{\max}$ and $\req$ for different numbers of servers. In Fig.~\ref{fig:ReqRmax}, we show the values of $r_{\max}$ and $\req$ as the number of servers increases. It can be seen that the minimum resetting rate needed to improve the region of convergence, $\req$, scales as a power law; that is, $\req\sim\alpha c^{\beta}$, where $\alpha>0$ and $\beta>2$. On the other hand, the resetting rate minimizing the zone of convergence scales linearly; that is, $r_{\max} \sim \gamma c$ with $\gamma\approx3.9293$. Therefore, as the number of servers increases, the resetting rate needed for stochastic resetting to improve the long-term behavior of the number of resources in the $G/M/c$ queue grows exponentially. In contrast, the resetting rate grows linearly, worsening convergence the most. This implies that, most of the time, the inclusion of stochastic resetting will be detrimental for the convergence zones of the $G/M/c$ queue.
\section{Discussion}
In this paper, we have developed a theoretical framework that combines results from multiple search-and-capture events with resetting in the interval, along with those from queueing theory, to study the convergence of the number of resources consumed by a target. One of our main findings is an explicit expression for a critical, convergence-inducing starting position, $x_{r}^{*}$, in terms of the remaining model parameters. We used this expression to explore the effects of stochastic resetting on the convergence of a $G/M/c$ queueing system in two settings: in the first, we fix the resetting rate and vary the interval length, and in the second, we fix the interval length and vary the resetting rate. In both cases, we found a transition in the effects of resetting, between widening and shrinking the convergence regions, marked by the threshold parameters $\Leq$ and $\req$. Moreover, we derived an implicit expression that enables us to numerically determine these threshold parameters. We were also able to determine the convergence rate of the critical, convergence-inducing starting position in the interval, $x_{r}^{*}$, to the one in the half line, $\xhl$, and exploit the fast convergence to derive an explicit expression for $\Leq$. Finally, we characterize the behavior of $\req$ as the number of servers increases and conclude that, as the number of servers increases, it becomes harder for stochastic resetting to expand the convergence regions in the parameter space.

There are several future directions for the exploration of queueing systems emerging from search processes with stochastic resetting. First, we could extend the model to a broader range of examples, including general non-Markovian service times (i.e., $G/G/c$ queueing systems), general resetting distributions, and non-Markovian search processes. It would also be of interest to establish analytical results for the asymptotic behavior of $x_{r}^{*}$ as $L\to\infty$ and extend the analysis to when $r\to\infty$. 

Another future direction is to consider the effects of stochastic resetting in the regions of convergence of several targets receiving resources from the same search process. Several authors have studied the impact of stochastic resetting for search processes with various targets \cite{pal_first_2017, belan_restart_2018, bressloff_target_2020}, and the convergence regions for a pair of targets receiving resources from the same search process were studied in Ref.~\cite{giral-barajas_stochastic_2025}; however, the combination of these two effects remains unexplored. Considering more than two targets underscores the need to understand the impact of resetting on the convergence of the number of resources in a 2D or 3D domain with a higher-dimensional search process.

\rev{A complementary direction is to study the effects of stochastic resetting in queues being fed by multiple interacting searchers. Systems with multiple searchers allow for additional resetting mechanisms, such as global resetting, in which all searchers are reset simultaneously, and independent resetting, in which each searcher resets individually \cite{nagar_stochastic_2023}. Recent work has examined the effects of these protocols on first-passage properties for systems of interacting and noninteracting particles \cite{biroli_critical_2023,biroli_exact_2024,vatash_many-body_2025,vilk_collective_2026,boyer_emerging_2026, lee_general_2026}. It would be of great interest to extending the present framework to such settings and study how interactions between searchers and the nature of the resetting protocol affect the arrival distribution and, in turn, the convergence of the associated queueing system}

\rev{Finally, it would} be interesting to study the steady-state distribution of the number of resources under heavy traffic, when the starting position approximates the critical value $x_{r}^{*}$ from the inside of the convergence zone and $\rho\uparrow 1$. Finally, note that within the context of queueing theory, we explore the effects of stochastic resetting on the inter-arrival time distribution by identifying the latter with the FPT statistics of a search process with resetting. An alternative perspective is to incorporate resetting into the queue length \cite{roy_queues_2024} and service times \cite{bonomo_mitigating_2022, bonomo_queues_2025}. These modifications allow for new nontrivial dynamics in the queue length. 
\begin{acknowledgments}
Jos\'{e} Giral-Barajas was supported by a Roth PhD scholarship funded by the Department of Mathematics at Imperial College London.
\end{acknowledgments}

\bigskip
\appendix
\section{\label{app:A}Starting position for the expediting of the search process with resetting}
From Eqs.~(\ref{eq:FPTDensity}) and (\ref{eq:LaplaceFlux}) we have that
\begin{equation}
	\widetilde{f}_{0}(x_{0},t)=\frac{\cosh\left[\sqrt{\frac{s}{D}}(L-x_{0})\right]}{\cosh{\left[\sqrt{\frac{s}{D}}L\right]}}\;,
	\label{eq:SurivalLaplace}
\end{equation}
and we can obtain the $n$th FPT moment directly from the Laplace transform of the FPT density
\begin{equation}
	T_{0}^{(n)}(x_{0}) := \mathbb{E}[\mathcal{T}_{0}(x_{0})^{n}] = -\frac{\partial^{n}}{\partial s^{n}}\widetilde{f}_{0}(x_{0},s)\Big|_{s=0}\;.
	\label{eq:FPTMomentsLaplace}
\end{equation}
Differentiating and taking the limit $s\to0$, we obtain
\begin{subequations}
\begin{eqnarray}
T_{0}(x_{0})&=&\frac{(2L-x_{0})x_{0}}{2D}\;, \label{eq:MFPT1}
\\
T_{0}^{(2)}(x_{0})&=&\frac{x_{0}^{4}-4x_{0}^{3}L+8x_{0}L^{3}}{12D^{2}}\;. \label{eq:MFPT2}
\end{eqnarray}
\end{subequations}
Now, given that $\sigma_{0}^{2}(x_{0})=T_{0}^{(2)}(x_{0})-T_{0}(x_{0})^{2}$, we can rewrite Eq.~(\ref{eq:CVCondition}) as
\begin{equation}
	\Phi(x_{0}):=T_{0}^{(2)}(x_{0}) - 2T_{0}(x_{0})^{2}  > 0\;.
	\label{eq:CVConditionAP1}
\end{equation}
Substituting Eqs.~(\ref{eq:MFPT1}) and (\ref{eq:MFPT2}) into this condition and simplifying the resulting expression, we obtain that $\Phi(x_{0})>0$ if and only if 
\begin{equation}
	\frac{-5x_{0}^{4}+20Lx_{0}^{3}-24L^{2}x_{0}^{2}+8L^{3}x_{0}}{12D^{2}} > 0\;.
	\label{eq:CVConditionAP2}
\end{equation}
Solving the left-hand side of this inequality for $x_{0}$ we obtain four roots
\begin{subequations}
\begin{eqnarray}
x_{1}^{*}&=&0\;, \label{eq:root1}
\\
x_{2}^{*}&=&\left(1-\frac{1}{\sqrt{5}}\right)L\;, \label{eq:root2}
\\
x_{3}^{*}&=&\left(1+\frac{1}{\sqrt{5}}\right)L\;, \label{eq:root3}
\\
x_{4}^{*}&=&2L\;. \label{eq:root4}
\end{eqnarray}
\end{subequations}

\begin{figure}[t!]
\includegraphics[width=\columnwidth]{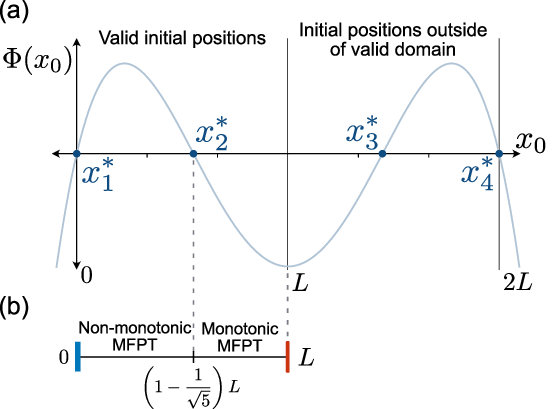}
\caption{\label{fig:roots} Roots for the polynomial $\Phi(x_{0})$ and subregion of $[0,L]$ for the initial positions of the search process ensuring a non-monotonic MFPT as a function of the resetting rate, $r$, and the possible expediting of the process.}
\end{figure}

With these roots, we are able to determine that $\Phi(x_{0})>0$ if and only if $x_{0}\in(0,[1-1/\sqrt{5}]L)$ or $x_{0}\in((1+1/\sqrt{5})L,2L)$, see Fig.~\ref{fig:roots}~(a). However, the only physically realistic domain for the initial position of the search process is $((1+1/\sqrt{5})L,2L)$, as shown in Fig.~\ref{fig:roots}~(b). Therefore, we finally obtain that the subregion of $[0,L]$ for which an initial position allows stochastic resetting to expedite the search process is
\begin{equation}
	\mathcal{X}_{0}=\left(0,\left[1-\frac{1}{\sqrt{5}}\right]\cdot L\right)\;.
	\label{eq:SubregionAP}
\end{equation}

\section{\label{app:B}Partial derivative of the critical starting position with resetting}
For the sake of completeness, we include the resulting expression for the partial derivative with respect to the resetting rate of the critical, convergence-inducing starting position with resetting:
\begin{widetext}
\begin{equation}
	\frac{\partial}{\partial r}x_{r}^{*}=\sqrt{\frac{D}{4r^{3}}} \left(\arccosh\left[\frac{c \mu  \cosh \left(L \sqrt{\frac{r}{D}}\right)}{c \mu -c \mu  r \tau +r}\right]-\frac{c \mu  r \left(\frac{L (c \mu -c \mu  r \tau +r) \sinh \left(\frac{L \sqrt{r}}{\sqrt{D}}\right)}{\sqrt{D} \sqrt{r}}+2 (c \mu  \tau -1) \cosh \left[L \sqrt{\frac{r}{D}}\right)\right]}{(c \mu -c \mu  r \tau +r)^2 \sqrt{\frac{c \mu  \cosh \left[L \sqrt{\frac{r}{D}}\right]}{c \mu -c \mu  r \tau +r}-1} \sqrt{\frac{c \mu  \cosh \left[L \sqrt{\frac{r}{D}}\right]}{c \mu -c \mu  r \tau +r}+1}}\right)\;.
	\label{eq:PartialDerivative}
\end{equation}
\end{widetext}

\providecommand{\noopsort}[1]{}\providecommand{\singleletter}[1]{#1}%

\end{document}